\documentclass[runningheads]{llncs}
\usepackage{graphicx}

\usepackage[T1]{fontenc}
\usepackage[utf8]{inputenc}
\usepackage[american]{babel}
\usepackage{textcomp}
\usepackage{csquotes}

\usepackage{booktabs}

\usepackage[hyphens]{url}
\usepackage[breaklinks=true, linkcolor=blue, urlcolor=blue, colorlinks=true]{hyperref}

\begin{document}
\title{A project-based course on software development for (engineering) research}
%
%
\author{Kyle E.~Niemeyer\inst{1}\orcidID{0000-0003-4425-7097}}
\authorrunning{K.E.~Niemeyer}
%
\institute{Oregon State University, Corvallis OR 97331, USA \\
\email{kyle.niemeyer@oregonstate.edu}} 
%
\maketitle              
\begin{abstract}
%
This paper describes the motivation and design of a 10-week graduate course that teaches
practices for developing research software; although offered by an engineering program,
the content applies broadly to any field of scientific research where
software may be developed. Topics taught in the course include local and remote version control,
licensing and copyright, structuring Python modules, testing and test coverage, continuous integration,
packaging and distribution, open science, software citation, and reproducibility basics, among others.
Lectures are supplemented by in-class activities and discussions, and all course material is shared openly
via GitHub. Coursework is heavily based on a single, term-long project where students individually develop
a software package targeted at their own research topic; all contributions must be submitted as pull requests
and reviewed\slash merged by other students. The course was initially offered in Spring 2018 with 17 students
enrolled, and will be taught again in Spring 2019.
\end{abstract}
\section{Motivation}

Nearly all research relies on software---even experimental---but researchers typically do not receive
training in best practices during graduate school in the same way as they do for experimental methods.
In fact, in two recent surveys the vast majority of academics confirmed that they use software and that
their research would be impractical without it: 90\%\slash 70\% of UK academics surveyed in 2014~\cite{Hettrick},
and 95\%\slash 63\% of US postdoctoral researchers surveyed in 2017~\cite{Nangia:2017aa}.
Computational science in particular depends on software and following good, evidence-based
practices when working with software and data.

For example, in the Mechanical Engineering graduate curriculum at Oregon State University,
the thermal-fluid sciences option (where I teach) requires a course on experimental measurement techniques,
but no analogous course on proper techniques on software development or computational science.
(We do require a course on numerical methods that focuses on solving differential equations,
but this does not extend to software development.) Instead, our program---as in most similar programs
around the world---assumes that such practices are trivial compared with the physical phenomena or
mathematical methods and\slash or can be self-taught. However, in the same way that appropriate
measurement techniques and statistical analysis of data are necessary for experimental (and computational)
research, good practices ensure reliability and correctness of research results obtained from software
developed for both computational (i.e., modeling-based) or experimental (i.e., analysis of results) research.

As the research community has recognized the importance of software and data skills, in recent years
Software Carpentry~\cite{Wilson2016} workshops have become a recognized avenue for graduate students and
postdocs (and the occasional faculty member) to learn necessary skills for working with Python,
the command line, and version-control systems. While these are essential skills for research,
researchers who go further to develop software require additional training.
This article describes a course aimed at filling this gap by teaching graduate-student
researchers practical software development skills, and also exposing them to topics related
to open science and reproducibility.

\section{Course design}

The course heavily relies and builds on the \textit{Effective Computation in Physics}
textbook by Scopatz and Huff~\cite{scopatz2015effective} (Chapters 10--22),
as well as recommendations by Wilson et al.~\cite{Wilson2017} and Jim\'{e}nez et al.~\cite{Jimnez2017}.
All materials for the course are openly available online via GitHub and shared under a Creative Commons
Attribution license; the online course syllabus provides links to each lecture~\cite{Niemeyer:2018aa}.
The course combines lectures, hosted on GitHub and presented using \texttt{reveal.js}~\cite{El-Hattab:2018aa},
with in-class activities and discussions, as summarized in Table~\ref{tab:schedule}.
Out-of-class work, described in Table~\ref{tab:assignments}, centers around a
software development project discussed in Sec.~\ref{sec:project}.

\begin{table}[htb]
\centering
\caption{Course schedule over ten weeks, with in-class activities.}
\label{tab:schedule}
\begin{tabular}{@{}ll@{}}
\toprule
Topic & In-class activity \\
\midrule
Getting started, and version control & Configure Git\\
Remote version control, licensing, and copyright & Create, clone, and fork repos \\
Structuring modules, and testing & Create basic structure of module \\
Test coverage, continuous integration, documentation & Configure Travis CI \\
Introduction to Julia (guest lecture) \\
Introduction to parallel programming & \\
Classes and objects (in Python) \\
Packaging and distributing your software & Create PyPI, Anaconda packages \\
Optimizing numerical code in Python \\
Working with files, command-line inputs in Python \\
Open science, software citation, reproducibility & Connect GitHub and Zenodo \\
Posters, presentations, and technical writing \\
Project presentations \\
\bottomrule
\end{tabular}
\end{table}

\begin{table}[htb]
\centering
\caption{Assignments, with week assigned.}
\label{tab:assignments}
\begin{tabular}{@{}lc@{}}
\toprule
Assignment & Week \\
\midrule
Join Gitter chat room and create GitHub profile & 1 \\
Project proposal & 1 \\
Choose open-source license & 3 \\
Create tests and submit as PR for review & 3 \\
Finish configuring Travis CI & 4 \\
Write comments and docstrings & 4 \\
Complete PyPI and\slash or Anaconda packages & 6 \\
Write report and make presentation & 7 \\
\bottomrule
\end{tabular}
\end{table}

\subsection{Course description and learning objectives}

The listed course description is
\begin{displayquote}
This course will advance students' understanding of topics related to computational
science and engineering, and advance their skills in applying techniques to solve
research problems using high-level programming languages. The course will build on
existing abilities in computer programming to cover topics related to computational
modeling and scientific software development. Students will gain experience in applying
available packages and libraries, as well as developing software to solve problems
related to their own research interests. Students will also gain experience in working
collaboratively and openly on scientific computing projects.
\end{displayquote}
By the end of the course, students should be able to
\begin{enumerate}
\item use high-level programming language to analyze and/or solve practical research problems;
\item apply principles of modern computational science and engineering, reproducibility,
and open science to their research;
\item evaluate, visualize, write about, and publish computational research results; and
\item develop and share an open-source research software package that solves a problem
in their research area.
\end{enumerate}
These are the formal student learning objectives for the course.

\subsection{Project}
\label{sec:project}

In lieu of standalone homework assignments, all assigned work contributes to a term-long project
where students develop a new software package targeted at their own research area.
The project initiates with a proposal that students submit via pull request to an open
repository on GitHub, and which the instructor merges upon approval (following any changes
requested). Then, students create a repository in the course organization for their software package,
and fork this to their own accounts. After this, students submit all project contributions as pull
requests to the upstream repository. Partners review these and either approve or request changes;
only after the code-review partner approves the contribution can the project's owner merge the
pull request.

\subsection{Methods of instruction}

The course is delivered using a combination of lectures, discussion, and in-class
interactive work. Lectures mostly exist as \texttt{reveal.js}~\cite{revealjs} presentations,
which are shared openly on a public-facing syllabus website~\cite{Niemeyer:2018aa}.
Lectures also use practical demonstrations of Python code, shown using either
IPython~\cite{PER-GRA:2007} or via Jupyter Notebooks~\cite{Kluyver:2016aa}.
Nearly all lectures also ask students to follow along on their own computer,
either executing example code or advancing their project packages.

\section{Results from first offering}

I offered the first iteration of this course in the Spring 2018 term, with the title
``Software Development for Engineering Research;'' while the course content is not
limited to engineering research, I offered the course out of the Mechanical Engineering
program with a targeted audience of graduate students in the College of Engineering.
17 students enrolled in the course, with all but one being graduate students; roughly
half were in the second year or later of their graduate programs.
Approximately 40\% of the students came from mechanical engineering (including thermal-fluid sciences
and design engineering), 35\% were from nuclear engineering, and the remaining came from robotics
and chemical engineering. Three quarters of the students had already taken a course on Python programming
for engineering applications, while the others had some self-taught Python programming skills.
Half expressed comfort working with the Unix command line, and the other half said they had used it
but were not as comfortable with command-line operations. None admitted to being command-line ninjas,
and none were completely unfamiliar.

The first offering of this 10-week course on software development for engineering research
completed successfully in June 2018, with all 17 students releasing the first version of their
software developed during the course. At least four of the software packages have been developed
further after the conclusion of the course, and at least one package is being prepared for
submission to the \textit{Journal of Open Source Software} (JOSS)~\cite{Smith2018}.

The projects covered a wide variety of topics, with functions including simulation,
experimental data analysis, and automation:
designing detonation tubes~\cite{Carter:2018aa},
using machine learning to extract features from nuclear physics simulations~\cite{Grechanuk:2018aa},
interfacing with an 8-channel digital pulse processor board~\cite{Mannino:2018aa},
simulating and analyzing the combustion engine of a Global Formula Racing
formula SAE vehicle~\cite{Kittelman:2018aa},
optimizing and analyzing wind-farm layouts~\cite{Miller:2018aa},
analyzing spin stabilization of solid rocket motors~\cite{Morse:2018aa},
a nodal quasi-diffusion solver for nuclear fission~\cite{Reynolds:2018aa},
agent-learning for autonomous path finding~\cite{Sripada:2018aa},
generating input for a Monte-Carlo radiation transport code~\cite{Stewart:2018aa},
calculating solar-energy terms based on location~\cite{Guymer:2018aa},
analyzing radioxenon spectra~\cite{Czyz:2018aa},
calculating deep-learning layers for multi-agent reinforcement learners~\cite{Brian:2018aa},
analyzing solvent extraction kinetics~\cite{Bettinardi:2018aa},
simulating rapid compression machine experiments~\cite{Behnoudfar:2018aa},
calibrating blackbody infrared cameras~\cite{Bean:2018aa}, and
simulating transient heat transfer in a microchannel with passive
temperature dependent flow control~\cite{Armatis:2018aa}.

Although the sample size is small, students rated the course well in their end-of-term evaluations:
they rated the course as a whole 5.3\slash 5.5 (mean\slash median) out of 6.0 and the instructor contribution
5.5\slash 5.8 (mean\slash median) out of 6.0. Multiple comments discussed the course favorably, and
that it should be taken by all students doing research involving software\slash programming.
Suggestions included clarifying expectations for students and proving more feedback; also, one student
felt the course was too advanced for their experience level.

\section{Conclusions}

This article describes a 10-week course given in Spring 2018 teaching
skills for developing research software; all lesson and assignment content is
available openly~\cite{Niemeyer:2018aa}.
This course will be offered again in Spring term 2019 (10 weeks, April--June).
Planned changes include incorporating more in-class activities in more of the topics,
and adding new topics such as peer code review and high-performance computing.

In addition, I am developing alternate versions of the course aimed at different lengths of time,
such as an afternoon tutorial session or day-long workshop. These lessons and modules will
be shared openly for the community to use, adapt, and extend.
Furthermore, while the course at Oregon State University is currently offered out of the
Mechanical Engineering program, it may fit better offered as an Engineering course or more broadly
in the Graduate Education program.

\section*{Acknowledgements}
This research was supported by the Better Scientific Software Fellowship, part of the Exascale
Computing Project (17-SC-20-SC), a collaborative effort of the U.S. Department of Energy Office of
Science and the National Nuclear Security Administration.

%
%
%
\bibliographystyle{splncs04}
\bibliography{refs}
%
%
%
%
%

\end{document}